\documentclass{article}

\usepackage{arxiv}

\usepackage[T1]{fontenc}    
\usepackage{hyperref}       
\usepackage{url}            
\usepackage{booktabs}       
\usepackage{amsfonts}       
\usepackage{nicefrac}       
\usepackage{microtype}      
\usepackage{lipsum}		
\usepackage{graphicx}
\usepackage[square,sort,comma,numbers]{natbib}
\usepackage{doi}

\usepackage{amsmath,newtxmath}
\DeclareMathOperator{\cat}{cat}
\DeclareMathOperator{\meanspec}{MeanSpec}
\DeclareMathOperator{\atan}{atan2d}

\title{Sliding-Window Normalization to Improve the Performance of Machine-Learning Models for Real-Time Motion Prediction Using Electromyography}

\author{ 
    Taichi Tanaka\\
	Science Technology of Innovation\\
	Nagaoka University of Technology\\
	Nigata, Japsan \\
	\texttt{taichi\_tanaka@stn.nagaokaut.ac.jp} \\
	\And
	Isao Nambu \\
	Department of Electrical Engineering\\
	Nagaoka University of Technology\\
	Nigata, Japsan\\
	\texttt{inambu@vos.nagaokaut.ac.jp}\\
	\And
	Yoshiko Maruyama \\
	Department of Production Systems Engineering\\
	National Institute of Technology, Hakodate College\\
	Hokkaido, Japan\\
	\texttt{yoshiko@hakodate-ct.ac.jp} \\
	\And
	Yasuhiro Wada \\
	Department of Electrical Engineering\\
	Nagaoka University of Technology\\
	Nigata, Japsan\\
	\texttt{ywada@vos.nagaokaut.ac.jp}\\
}

\renewcommand{\shorttitle}{Sliding-Window Normalization to Improve the Performance of Machine-Learning Models for Real-Time Motion Prediction Using Electromyography}

\hypersetup{
pdftitle={Sliding-Window Normalization to Improve the Performance of Machine-Learning Models for Real-Time Motion Prediction Using Electromyography},
pdfsubject={q-bio.NC, q-bio.QM},
pdfauthor={taichi tanaka},
pdfkeywords={Electromyography, EMG, Z-score, Signal Normalization, Machine-Learning, Classification Model},
}

\begin{document}
\maketitle

\begin{abstract}
	Many researchers have used machine learning models to control artificial hands, walking aids, assistance suits, etc., using the biological signal of electromyography (EMG). The use of such devices requires high classification accuracy of machine learning models. One method for improving the classification performance of machine learning models is normalization, such as z-score. However, normalization is not used in most EMG-based motion prediction studies, because of the need for calibration and fluctuation of reference value for calibration (cannot re-use). Therefore, in this study, we proposed a normalization method that combines sliding-window analysis and z-score normalization, that can be implemented in real-time processing without need for calibration. The effectiveness of this normalization method was confirmed by conducting a single-joint movement experiment of the elbow and predicting its rest, flexion, and extension movements from the EMG signal. The proposed normalization method achieved a mean accuracy of 64.6\%, an improvement of 15.0\% compared to the non-normalization case (mean of 49.8\%). Furthermore, to improve practical applications, recent research has focused on reducing the user data required for model learning and improving classification performance in models learned from other people’s data. Therefore, we investigated the classification performance of the model learned from other’s data. Results showed a mean accuracy of 56.5\% when the proposed method was applied, an improvement of 11.1\% compared to the non-normalization case (mean of 44.1\%). These two results showed the effectiveness of the simple and easy-to-implement method, and that the classification performance of the machine learning model could be improved.
\end{abstract}



\section{Introduction}
Electromyography (EMG) is a biological signal whose amplitude fluctuates when exercising or contracting muscles. Many researchers have used this property to research and develop devices that are aimed at expanding and recovering human motor function [1-5]. Due to its easy design, machine learning models have been used in many studies including motion control for artificial hands and gesture recognition using classifiers, and control of walking aids and assistance suits by predicting joint angles, joint angular velocities, or joint torque using regressors [2-5]. Linear models such as logistic regression and support vector machines were first used around 2000, with an emphasis on improving classification performance by the feature extraction method [5-8]. However, as classification performance significantly improved with the development of deep learning [9] that occurred in 2012, research was also conducted to improve classification performance by changing the configuration of the deep neural network [10-12].\\

 Data normalization is one of the methods for improving the classification performance of machine learning models and is used in fields such as imaging and biometrics [13-15]. Methods that are often used include min-max normalization [15, 16], which normalizes the value range of the dataset from 0 to 1; and z-score, which normalizes the dataset mean to 0 and standard deviation to 1 [15, 17]. Even in the field of EMG, the classification performance of models is improved by normalizing signals and features with z-score and min-max normalization [16-20]. Although the EMG fields use normalization such as maximum voluntary contraction (MVC) [21] or maximum voluntary isometric contraction (MVIC) [22] to enable motor analysis and motor performance evaluation between muscles and subjects, normalization is hardly used in EMG-based motion prediction research. It is thought that there are two possible reasons why normalization is rarely used in motion prediction research. The first reason is the need to measure the reference value (e.g., min, max, mean, or standard deviation of each EMG channel) to carry out the calibration. It can take 30 s –3 min to use the application, depending on the measurement method. The second reason is that reference values such as the max and mean of each EMG channel fluctuate due to various factors such as muscle fatigue, electrode position, and fluctuations in skin impedance [23-27]. Therefore, a reference value, once measured, cannot be re-used. This reduces the practical applications of normalization and makes it unsuitable for real-time processing (i.e., online processing). Therefore, we aimed to devise a normalization method that does not require calibration (i.e., measurement of reference values) and that is suited for real-time processing to enable normalization to be used as a means of improving the prediction performance of machine learning models.\\
 
We propose a normalization method that uses the sliding-window [28] and z-score normalization [15, 17] shown in Section II.B. The z-score is a simple normalization method that sets the dataset mean to 0 and standard deviation to 1. Compared to min-max normalization, which uses the minimum and maximum values in the entire dataset, the z-score uses the mean and standard deviation of the entire data, making it less susceptible to outliers. However, the z-score is usually not suitable for real-time processing because the entire dataset needs to be used for normalization, which incurs a time delay. Therefore, we considered combining the sliding-window analysis (SWA) that is used for signal analysis with time-varying parameter analysis. SWA involves analyses that use the signal within a specified window length. It is thought that using a signal of a sufficient length can achieve the same effect as the z-score that uses the entire dataset. \\

In recent years, research has focused toward enabling other people’s machine learning models to exhibit the same classification performance as machine learning models trained from their own data (i.e., improving versatility among users) [29]. Studies that solve the problem of requiring individually specialized machine learning models by measuring a large amount of data for each user due to individual differences in myoelectric amplitudes have been reported. Methods have been proposed to reduce the required amount of own data by using other people’s data with domain adaptation, which technology enables the use of models that were trained in different datasets, even in datasets with different data attributes [29–31]. Such methods include Geodesic Flow Kernel (GFK) [32], correlation alignment (CORAL) [33] and transfer component analysis (TCA) [34] which conducts motion prediction using a machine learning model trained from a different dataset after projecting one’s own data onto the data space, and domain adversarial neural networks (DANN) [35] which is a kind of deep learning method which trains the model to extract common features between different datasets.\\

The proposed method normalizes the standard deviation of myoelectric amplitude with individual differences, so it is thought that the influence of individual differences in myoelectric amplitude can be reduced, and the classification performance in the model learned from a different subject’s dataset can be improved. Compared to previous research such as DANN, the proposed method trains machine learning models using data other subject than one’s own data, so it is superior in that the models do not need to be trained for each user.\\
\section{Methods}
\subsection{Proposed Sliding-Window Normalization}
We propose a normalization method using sliding-window analysis (SWA) and z-score to improve the classification performance of the machine learning model and versatility between users (i.e., exhibiting the same classification performance as the own machine learning model in the other’s machine learning model). SWA is used for signal analysis and time-varying parameter analysis using the signal within a specified window length [28]. SWA enables time series analysis by sliding the window so that when a new sample is obtained, the sliding-window replaces the oldest sample with the new sample. The z-score is a kind of normalization method that is used to improve the classification performance of models in machine learning. The features are normalized by setting the feature mean to 0 and the standard deviation to 1 [15, 17].\\

The proposed method is a combination of these two concepts and is called sliding-window normalization (SWN). As shown in Eq. (1), the mean and standard deviation of the samples in the sliding-window are set to 0 and 1, respectively.\\
\begin{eqnarray}
{SWN\ EMG}_{t,\ n-t+L_{norm}}=({EMG}_n-m_t)/\sigma_t\\
(t-L_{norm}<n\le t) \nonumber \\
m_t=1/L_{norm}\sum_{i=0}^{L_{norm}-1}{EMG}_{t-i} \nonumber \\ \sigma_t=\sqrt{1/L_{norm}\sum_{i=0}^{L_{norm}-1}({EMG}_{t-i}-m_t)^2} \nonumber
\end{eqnarray}
Here, \begin{math}t\end{math} is the current discrete time, \begin{math}L_{norm}\end{math} is the sliding window length, \begin{math}n\end{math} is the discrete time number in the sliding window, \begin{math}EMG_{i}\end{math} is the \begin{math}i^{th}\end{math}  processed EMG, \begin{math}{SWN EMG}_{t,\ n-t+L_{norm}}\end{math} is the myoelectric signal to which the \begin{math}n-t+L_{norm}^{th}\end{math} proposed method (SWN) is applied at the \begin{math}t^{th}\end{math}, and \begin{math}m_{t}\end{math} and \begin{math}\sigma_{t}\end{math} are the myoelectric mean and standard deviation on the \begin{math}t^{th}\end{math} sliding window. We used the “mean” and “std” functions in numpy in Python.\\
This normalization method has the following advantages:
\begin{itemize}
\item No need for calibration.
\item Real-time classification and regression are possible because the processing is simple and the calculation time is short.
\item It can be combined with methods that improve versatility between users.
\end{itemize}

\subsection{Evaluation Method}
This paper evaluates three types of items. The first is the improvement of the classification performance of machine learning models when the proposed method (SWN) is applied (Section II.B.1), the second is the improvement of versatility between users of machine learning models when the proposed method (SWN) is applied (Section II.B.2), and the third is the improvement of the classification performance of the machine learning model when the number of subjects of the model that was trained with different data is increased by applying the proposed method (SWN) (Section II.B.3).\\

Two types of machine learning models need to be trained. The first is the model trained with one’s own data (model type of OWN). The second is the model trained with another person’s data (model type of OTHER). Sections II.B.1 and II.B.2 used models OWN and OTHER. Section II.B.3 used only OTHER. Performance (OWN) involved dividing the data into training and testing datasets and calculating the performance using the model trained with one’s own training data and own test data. Performance (OTHER) involved calculating the performance using the model trained with another person’s training data and one’s own test data. The training data and test data was created by randomly dividing them into a 1:1 ratio every 10 consecutive trials.\\

\subsubsection{Normalization Evaluation}
The evaluation of model classification performance improvement by the proposed method (SWN) was conducted by comparing the “classification performance in the model with normalization (OWN or OTHER)” and “classification performance of the model without normalization (OWN or OTHER)”.\\

Improvements in model classification performance due to the proposed model will be indicated by higher performance and lower standard deviation in performance. We consider  the model classification performance improved and the research objective achieved when the performance (OWN or OTHER) with SWN applied is equal to or greater than the performance (OWN or OTHER) with no SWN applied (no normalization).\\
\subsubsection{Versatility Evaluation}

The evaluation of versatility between users was conducted by comparing the “classification performance in the model trained with one’s own data (OWN)” with the “classification performance in the model trained with another person’s data (OTHER)”.\\

Better user’s versatility is indicted by higher performance and lower standard deviation. Versatility between users is considered improved and the research objective achieved when the performance (OTHER) with SWN applied is equal to or greater than the performance (OWN) with no SWN applied (no normalization).\\
\subsubsection{Evaluation on SWN Increased Number of Subject to Train Model}

We investigated whether the classification performance of the model could be improved by increasing the number of subjects used for learning the model. We compared a model trained with nine subjects (OTHER) with a model trained with one subject (OTHER). The model trained with nine subjects (OTHER) was considered good if its performance was high and its performance had a low standard deviation.\\
\subsubsection{Evaluation Index}

The accuracy shown in Eq. (2) was used as the evaluation index for the classification performance of the machine learning model. Accuracy is an evaluation index that can simply compare results with multiple targets.\\
\begin{eqnarray}
Accuracy= \frac{Success\ Predictons}{Success\ Predictons+Failure\ Predictons\ }
\end{eqnarray}

The Wilcoxon rank-sum test was used for significance tests. The significance level was set for the P-value less than 0.05. The “ranksums” function in scipy.states in Python was used for implementation. The “multipletests” function in statemodels.sandbox.stats.multicomp in Python was used for multiple comparisons. We used Bonferroni correction as the correction method for the P-value.\\
\subsubsection{Machine-Learning Model}
We chose multi-class logistic regression for the machine learning model, which allows multi-class classification. The “LogisticRegression” function in scikit-learn in Python was used for implementation. The parameters were as follows: penalty=“none”, class\_weight=“balanced”, and max\_iter=6000. The number of models trained was calculated by \begin{math}{}_{number of subject} C_{number of subject to train}\end{math}.

\subsection{EMG Proccesing}
Before training the machine learning model, the measured EMG underwent pre-processing, normalization, feature extraction, and decimation.\\

Pre-processing involved the application of a low-pass Butterworth filter (3rd-order, 500 Hz), decimation (2000 → 500 Hz), and a high-pass Butterworth filter (3rd order, 30 Hz). We used the scipy.signal “butter” function and “sosfilt” in Python for implementation. Normalization involved the application of either SWN or no normalization (i.e., None). The normalization window length was set at between 100 and 500 ms, with 100-ms intervals. In the case of “None”, the obtained features did not change even when the normalization window length was changed.\\

Feature extraction involved the calculation of the following five features, for which high classification performance was obtained in previous studies: mean absolute value: MAV (Eq. (3)) [6], mean waveform length: MWL (Eq. (4)) [7], and difference root mean square: DRMS (Eq. (5)) [7] as time-dimension features, and shot-time Fourier transform: STFT [5] and stationary wavelet transform: SWT [8] as frequency-dimension features. STFT involved averaging in the 1–70 Hz (low component), 60–100 Hz (middle component), and 100–250 Hz (high component) ranges and concatenating them (Eq. (6)). SWT involved time-frequency conversion using Daubechies wavelet 2 (db2) as the mother wavelet and taking the absolute mean of the wavelet coefficient of level 3 frequency (cD3) as the feature.\\
\begin{eqnarray}
{MAV}_t=1/L_{feature}\sum_{n=0}^{L_{feature}-1}|{EMG}_{t-n}|
\end{eqnarray}
\begin{eqnarray}
{MWL}_t=\frac{1}{L_{feature}-1}\sum_{n=0}^{L_{feature}-2}|{EMG}_{t-n}-{EMG}_{t-n-1}|
\end{eqnarray}
\begin{eqnarray}
{DRMS}_t=\sqrt{\frac{1}{L_{feature}-1}\sum_{n=0}^{L_{feature}-2}{({EMG}_{t-n}-{EMG}_{t-n-1})}^2}
\end{eqnarray}
\begin{eqnarray}
{STFT}_t=\rm cat(Low,Mid,Hig)\\
Low=\frac{1}{{bin}_{low}}\sum_{freq=1}^{70}{{\rm MeanSpec}_{freq}({EMG}_{(t-L+1)-t})}\nonumber\\
Mid=\frac{1}{{bin}_{mid}}\sum_{freq=60}^{100}{{ \rm MeanSpec}_{freq}({EMG}_{(t-L+1)-t})}\nonumber\\
Hig=\frac{1}{{bin}_{hig}}\sum_{freq=100}^{250}{{ \rm MeanSpec}_{freq}({EMG}_{(t-L+1)-t})}\nonumber
\end{eqnarray}
Here, \begin{math}t\end{math} is the current discrete time, \begin{math}L_{feature}\end{math} is the window length of feature extraction, \begin{math}\cat(\cdot)\end{math} is the concatenation function, \begin{math}freq\end{math} is the frequency, \begin{math}\meanspec(\cdot)\end{math} is the function that outputs the spectrogram averaged in the time direction, and \begin{math}bin\end{math} is the number of discrete frequencies in each of the low / middle / high frequencies. A Hanning window with a window length of 64 samples was used for the STFT window function. The functions in scipy.signal in Python were used for implementation. SWT is a method that improves the position invariance, which was a problem of wavelet transforms (WT), and the same mother wavelet as in WT can be used. The “swt” function in the pywt module in Python was used for implementation. The window length of feature extraction was set between 100 and 500 ms, with 100 ms intervals.

Finally, decimation involved reducing the sampling rate of the features from 500 Hz to 20 Hz to reduce the amount of data and shorten the training time of the machine learning model.\\

\subsection{WEIGHTING AND DIVIDING WINDOW FUNCTION}
We proposed a feature extraction method that can be used in combination with SWN to improve the classification performance of the models. The proposed feature extraction method is the weighting window function that weights the sample in the window shown in Fig. 1 in the time direction and the dividing window function that divides the window shown in Fig. 2 for feature extraction. Both the normalization and feature extraction window length were set between 100 to 500 ms, with 100-ms intervals.\\
\begin{figure}[tbhp]
  \begin{minipage}[b]{0.45\linewidth}
    \centering
    \includegraphics[keepaspectratio, width=8cm]{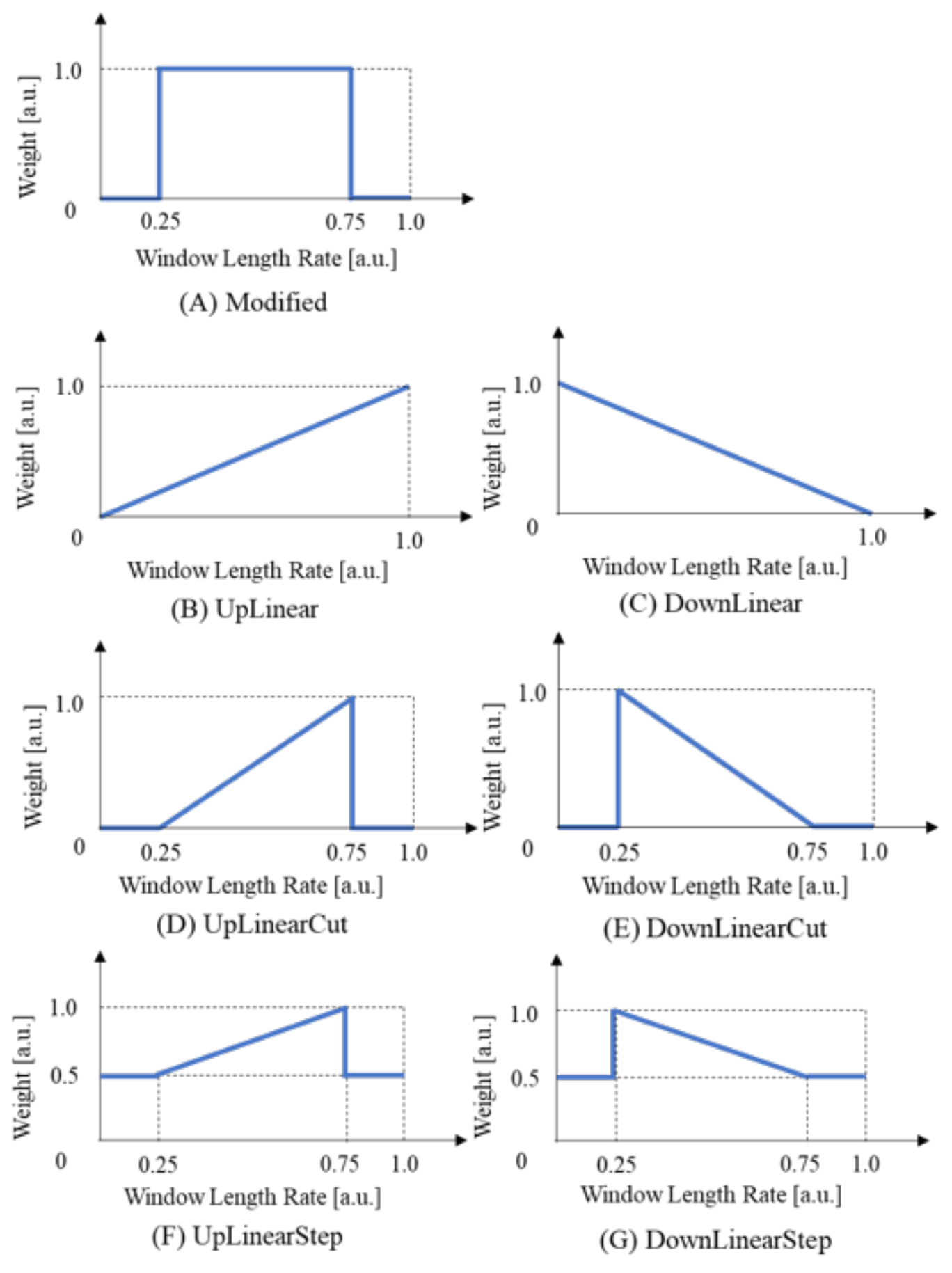}
    \caption{Weighting Window Functions.}
  \end{minipage}
  \begin{minipage}[b]{0.45\linewidth}
    \centering
    \includegraphics[keepaspectratio, width=8cm]{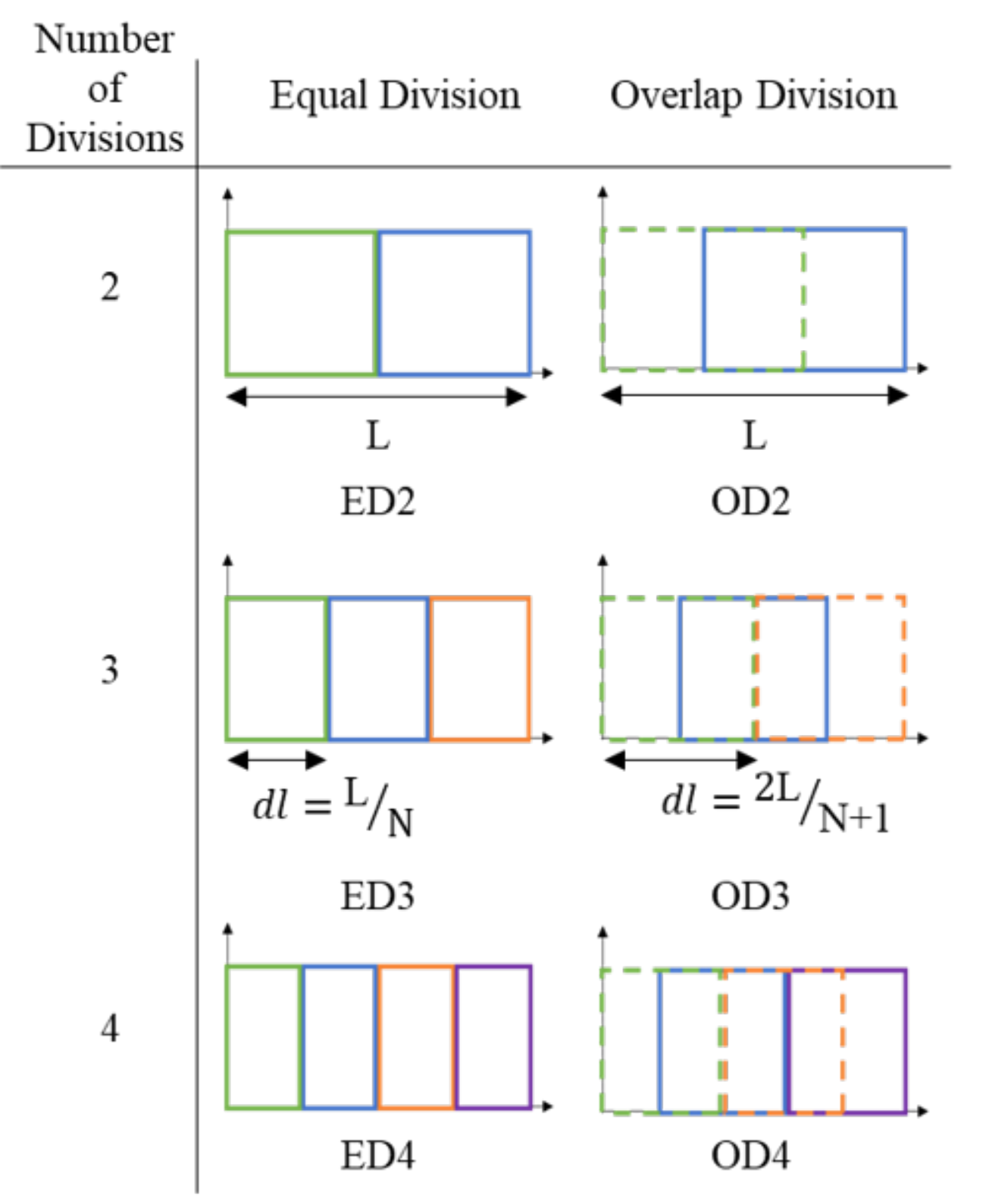}
    \caption{Dividing Window Functions. L: Window length, dl: Divided window length, N: Number of divisions.}
  \end{minipage}
\end{figure}

\subsubsection{Weighting Window Function}
We proposed seven types of weighting window functions that resulted in significant changes in EMG with respect to time, as shown in Fig. 1. Furthermore, we proposed the following methods of assembling paired window functions to further extract the change in EMG with respect to time: UpDownLinear ((B) UpLinear + (C) DownLinear), UpDownLinearCut ((D) UpLinearCut + (E) DownLinearCut), and UpDownLinearStep ((F) UpLinearStep + (G) DownLinearStep). Feature extraction was conducted using a total of 10 types of weighting window functions.\\

MAV, MWL, and DRMS applied the window function to the normalized EMG. STFT and SWT involved determining the absolute mean in the time direction by applying the window function after extracting the features of the STFT and SWT functions.\\
\subsubsection{Dividing Window Function}
We conducted six types of feature extraction, with two types of dividing window functions (Equal Division and Overlap Division) and three types of window division numbers (2, 3, 4), as shown in Fig. 2. A larger window division number shortens the divided signal length. Equal Division divides the input signal equally. Overlap Division divides the input signal so that half the divided signals overlap.\\

All five features were applied to the EMG after normalization. The MAV, MWL, and DRMS set the range of feature extraction window length between 100 and 500 ms, but the STFT and SWT cannot apply the window length range from 100 ms due to the specifications of the feature extraction method. STFT and SWT were applied only when the divided window length was over 100 ms.\\

\begin{figure}[tbhp]
\centering
\includegraphics[keepaspectratio, width=8cm]{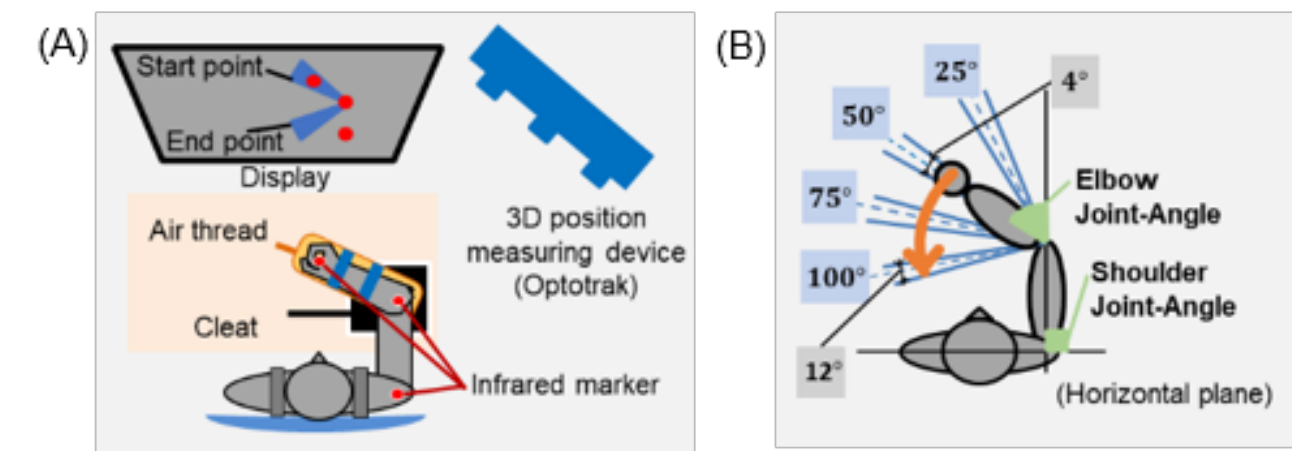}
\caption{Experiment Condition: (A) state, (B) task.}
\end{figure}

\subsection{DATA ACQUISITION}
\subsubsection{Subjects}
The ethics board of Nagaoka University of Technology approved this study according to the Declaration of Helsinki. The subjects were ten right-handed 22- to 23-year-old men. The subjects were informed about the experiment in advance and consented to participate in the experiment.\\
\subsubsection{Experiment}
The positions of the hands, elbows, and shoulders, and the EMG of the forearm and upper arm muscles were measured as in the experimental environment shown in Fig. 3(A). Subjects performed 12 types of elbow single-joint movements with four different start points and end points as tasks (Fig. 3(B)). A task involves moving from one of the four points (start point) to one of the other three points (end point). Each trial consisted of pre-rest (2.0 s), task (2.5 s), and post-rest (0.1 s); 36 trials (12 movements x 3) were conducted in one session, for a total of ten sessions (i.e., 360 trials). The tasks were randomly selected for each session. The following four rules were also set as the success conditions for the tasks.\\
\begin{enumerate}
\item No exercise during the rest period. Elbow joint angular velocity does not exceed 2.0 deg./s during the rest period.
\item End the task during the task period. End the task between 0–2.5 s.
\item Place the elbow joint angle at the start point ($\pm2.0$ deg.) during the rest period and at the end point ($\pm6.0$ deg.) at the end of the task.
\item Place the shoulder and elbow joints within 3 cm of the initial position between the pre-rest and post-rest.
\end{enumerate}

The position data were measured at three locations, namely, the hand, elbow, and shoulder, using Optotrack Certus, (NDI Inc., Waterloo, Canada, sampling rate: 500 Hz). The EMG was measured at the biceps brachii (×4), brachialis (×1), brachioradialis (×1), anconeus (×1), triceps brachii (outside) (×2), 	triceps brachii (long head) (×2), and extensor carpi radialis longus (×1), totaling 12 locations, by using Trigno Lab Avanti (Delsys, Natick, MA, USA, sampling rate: 2000 Hz).\\

\subsection{Position Processing}
Position processing consisted of noise reduction, work space → joint angle space conversion, elbow joint angular velocity conversion, coding, and decimation in order to obtain the target (rest, flexion, and extension of elbow joint movement) from the positions of the hands, elbows, and shoulders obtained in the subject experiment.\\

For noise reduction we applied a zero-phase low-pass Butterworth filter (2nd-order 20 Hz). The Python scipy.signal “butter” and “sosfiltfilt” functions were used.\\

Work space → joint angle space conversion involved the conversion of the positions of the hand, elbow, and shoulder to the elbow and shoulder joint angles using Eq. (7).\\
\begin{eqnarray}
\theta_{sld}=\atan(a,b)-\atan(a2+b2-c2,c)\\
\theta_{elb}=\atan(a2+b2-c2,c)+\atan(a2+b2-d2,d)\nonumber\\     
a=y_{hand}-y_{sld}\nonumber\\
b=x_{hand}-x_{sld}\nonumber\\
c=(a^2+b^2+{L_{sld}}^2-{L_{elb}}^2)/2L_{sld}\nonumber\\
d=(a^2+b^2-{L_{sld}}^2+{L_{elb}}^2)/2L_{elb}\nonumber
\end{eqnarray}
Here, \begin{math}\atan(y, x)\end{math} is the function that calculates the angle [deg.] from the two-dimensional coordinate position, \begin{math}x_{hand}\end{math} and \begin{math}y_{hand}\end{math} are the hand position [m], \begin{math}x_{sld}\end{math} and \begin{math}y_{sld}\end{math} are the shoulder position [m], and \begin{math}L_{sld}\end{math} and \begin{math}L_{elb}\end{math} are the upper arm and forearm length [m].
Elbow joint angular velocity conversion involved the conversion of the joint angle to the joint angular velocity using Eq. (8).\\
\begin{eqnarray}
{\dot{\theta}}_{elb,t}=(\theta_{elb,\ t+1}-\theta_{elb,\ t})f_s
\end{eqnarray}
Here, \begin{math}{\dot{\theta}}_{elb,t}\end{math} is the elbow joint angular velocity at the discrete time \begin{math}t\end{math}, and \begin{math}f_s\end{math} is the sampling frequency.

Coding involved the conversion of the elbow joint angular velocity to the target using Eq. (9).\\

This target was used as the teacher data for model training.\\
\begin{eqnarray}
{\rm target}_t=
\begin{cases}
{flexion \ \ (\theta_{elb,t} \geq 2.0 [deg./s])}\\
{extension \ \ (\theta_{elb,t} \leq -2.0 [deg./s])}\\
{rest \ \ (otherwise)}
\end{cases}
\end{eqnarray}

\section{Results}
Prior to evaluating the classification performance (Section III.B) and versatility between users (Section III.C) of the machine learning model by the proposed SWN method, we investigated the effects of window length for feature extraction and normalization (Section III.A). Afterwards, we investigated the effect of the number of subjects used in model OTHER (Section III.C). Finally, we investigated feature extraction methods (Section III.D) to further improve the accuracy. The chance level of accuracy in all results was 33.3\%.\\
\subsection{Effect of Window Length}
In model OWN, we investigated the effect of changing window length for feature extraction and normalization on accuracy. \\

First, the effect of window length for feature extraction on accuracy was investigated. Fig. 4 shows the results of changing the window length for feature extraction between 100 and 500 ms in 100-ms intervals and comparing the normalized (SWN with the window length fixed at 500 ms) and non-normalized (None) cases for the five types of features. Fig. 4 shows that applying SWN improved accuracy as the window length for feature extraction increased. In contrast, with no normalization, the accuracy decreased as the window length for feature extraction decreased. We surmise that applying SWN improves the classification performance of the model by lengthening the feature extraction window.\\

Next, the effect of window length for normalization on accuracy was investigated. We changed the window length for normalization between 100 and 500 ms at 100 ms intervals, and the window length for feature extraction was fixed at 500 ms. Fig. 5 shows the results of calculating with all five feature types. The accuracy did not change significantly even when the window length for normalization was changed. We suggest that the window length for normalization should be selected within the range of 100 to 500 ms, with the window length that maximizes accuracy being selected.\\

We also investigated whether there was any synergy between normalization and feature extraction window length, but no synergistic effects were observed.\\


\begin{figure}[tbhp]
  \begin{minipage}[b]{0.45\linewidth}
    \centering
    \includegraphics[keepaspectratio, width=7cm]{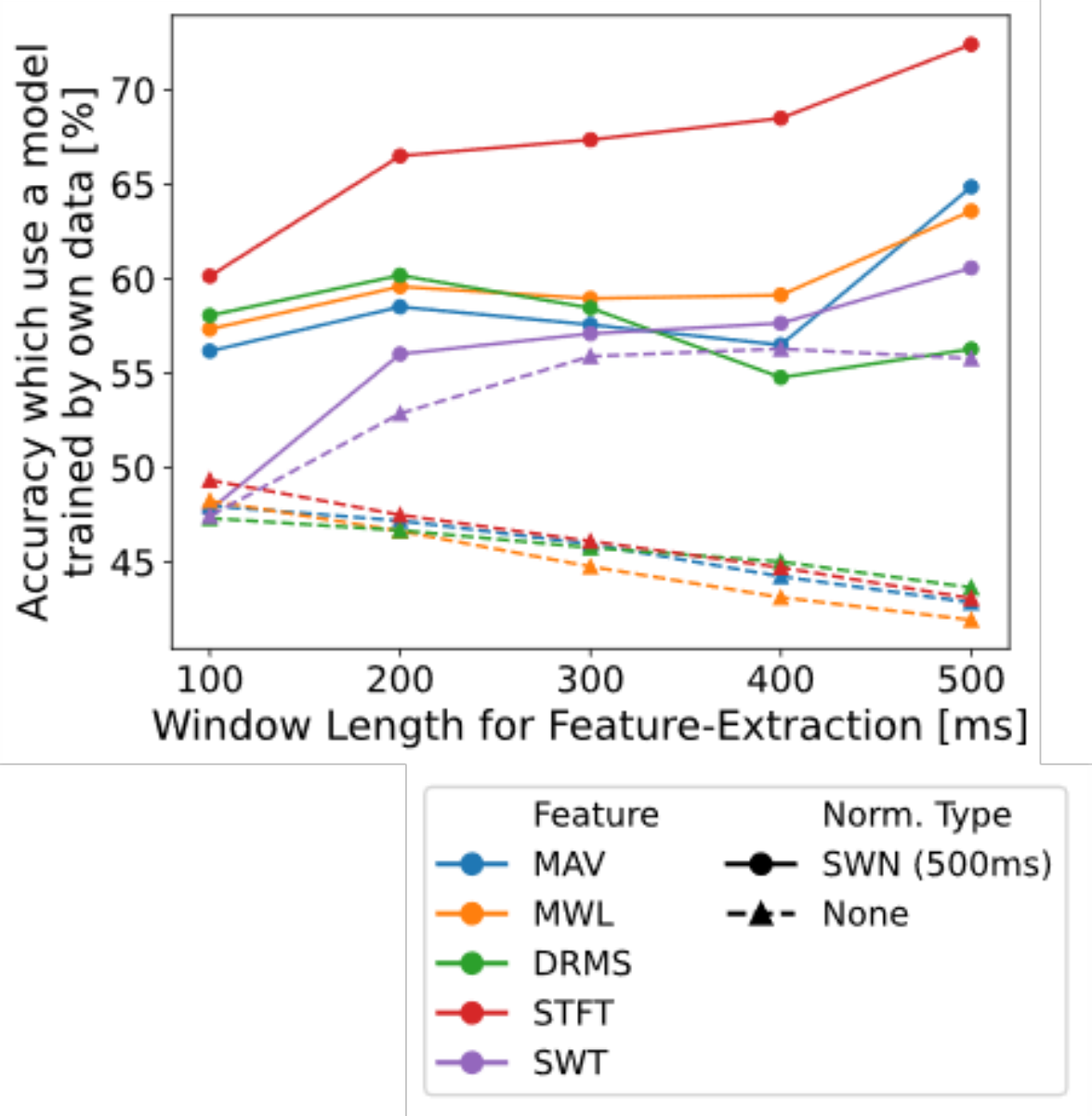}
    \caption{Effect of window length on feature-extraction (OWN).  Window length for SWN is fixed at 500 ms. Horizontal axis indicates window-length for feature-extraction and vertical axis indicates accuracy using a model trained by data from own subject. Each color line shows results for the type of feature-extraction. The solid line shows the results with the normalization (SWN which window length is fixed 500 ms), and the dashed line shows the results without the normalization (None).}
  \end{minipage}
  \begin{minipage}[b]{0.45\linewidth}
    \centering
    \includegraphics[keepaspectratio, width=7cm]{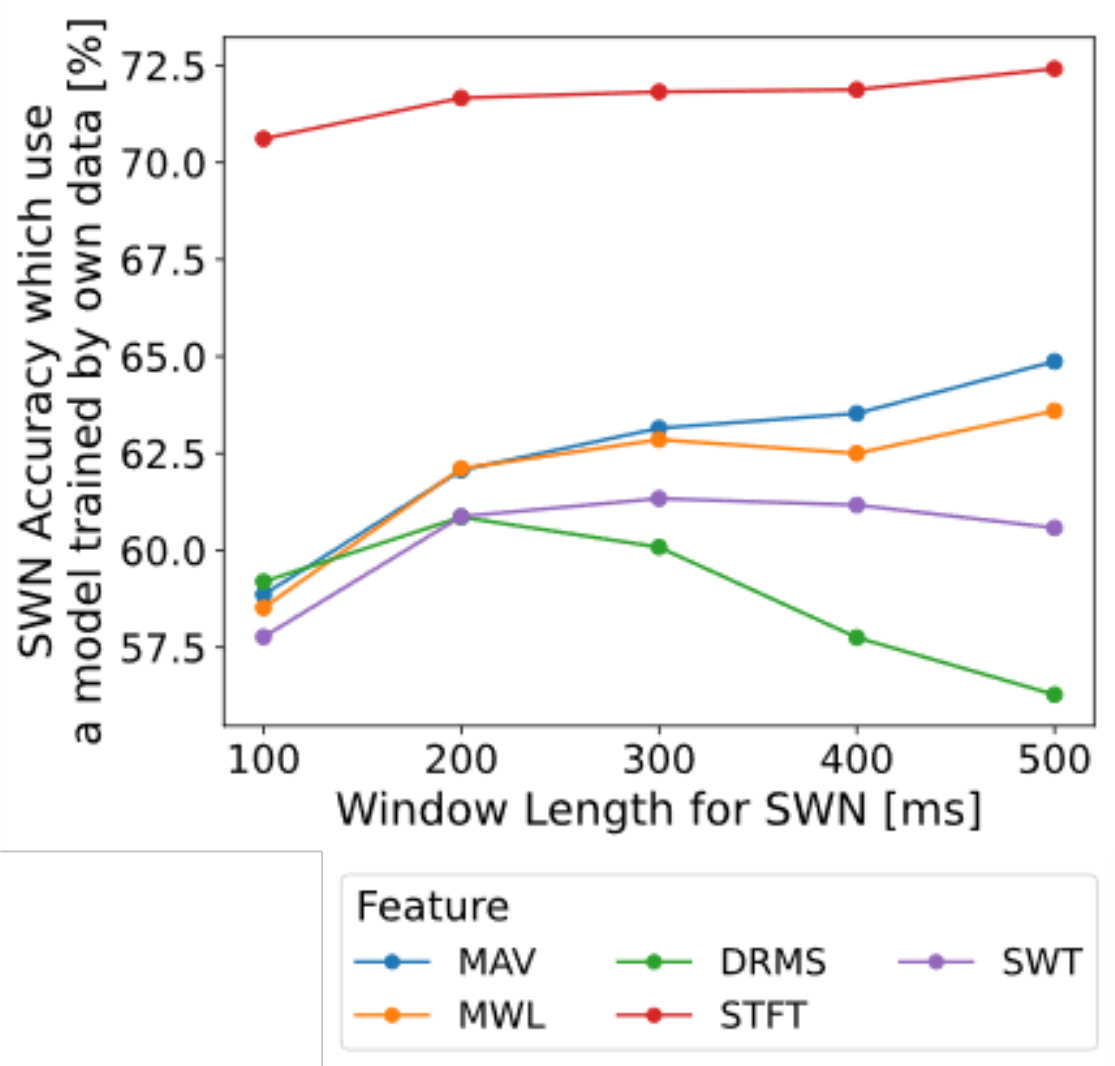}
    \caption{The effect of the window length for normalization (OWN). The window length for feature-extraction is fixed at 500 ms. Horizontal axis indicates window-length for normalization and vertical axis indicates accuracy using a model trained by data from own subject.}
  \end{minipage}
\end{figure}

\subsection{Normalization Comparison}
We investigated whether the proposed method (SWN) would improve the classification performance of the models. Research in recent years has been conducted to reduce the pre-data measurement of each user by enabling others’ machine learning models to exhibit the same classification performance as one’s own model (i.e., improving versatility between users). Therefore, in this study, we compared the accuracy of normalization (SWN) and non-normalization (None) between a model learned from one’s own data (OWN) and a model learned from other subjects’ data (OTHER). The window lengths for normalization and feature extraction were changed between 100 and 500 ms in 100-ms intervals, and the maximum accuracy was compared. The number of subjects used when training model OTHER was set to nine people. Fig. 6 shows the results.\\

We compared the accuracy of the model (OWN or OTHER) with (SWN) or without (None) normalization in Fig. 6. A comparison between SWN\_OWN (mean accuracy: 64.6\%, mean standard deviation of accuracy: 3.3\%, blue bar) and None\_OWN (mean accuracy: 49.8\%, mean standard deviation of accuracy: 9.1\%, orange bar) shows that the mean accuracy of SWN\_OWN increased by 15.0\% and its mean standard deviation of accuracy decreased by 5.8\%. These results show that the proposed method (SWN) was effective when using the machine learning model that was trained from one’s own data. Furthermore, a comparison between SWN\_OTHER (mean accuracy: 56.5\%, mean standard deviation of accuracy: 4.7\%, blue shaded bar) and None\_OTHER (mean accuracy: 44.1\%, mean standard deviation of accuracy: 12.0\%, orange shaded bar) shows that the mean accuracy of SWN\_OTHER increased by 11.1\% and its mean standard deviation of accuracy decreased by 6.5\%. These results show that the proposed method (SWN) was effective even when using other’s machine learning models. These two results show the effectiveness of the proposed method.\\

A comparison of the accuracy of each feature shows that STFT (accuracy: 72.4\%) was highest. The other accuracy values for MAV (64.9\%), MWL (63.6\%), DRMS (60.9\%), and SWT (61.3\%) were not much different.\\
\begin{figure}[tbhp]
\centering
\includegraphics[keepaspectratio, width=7cm]{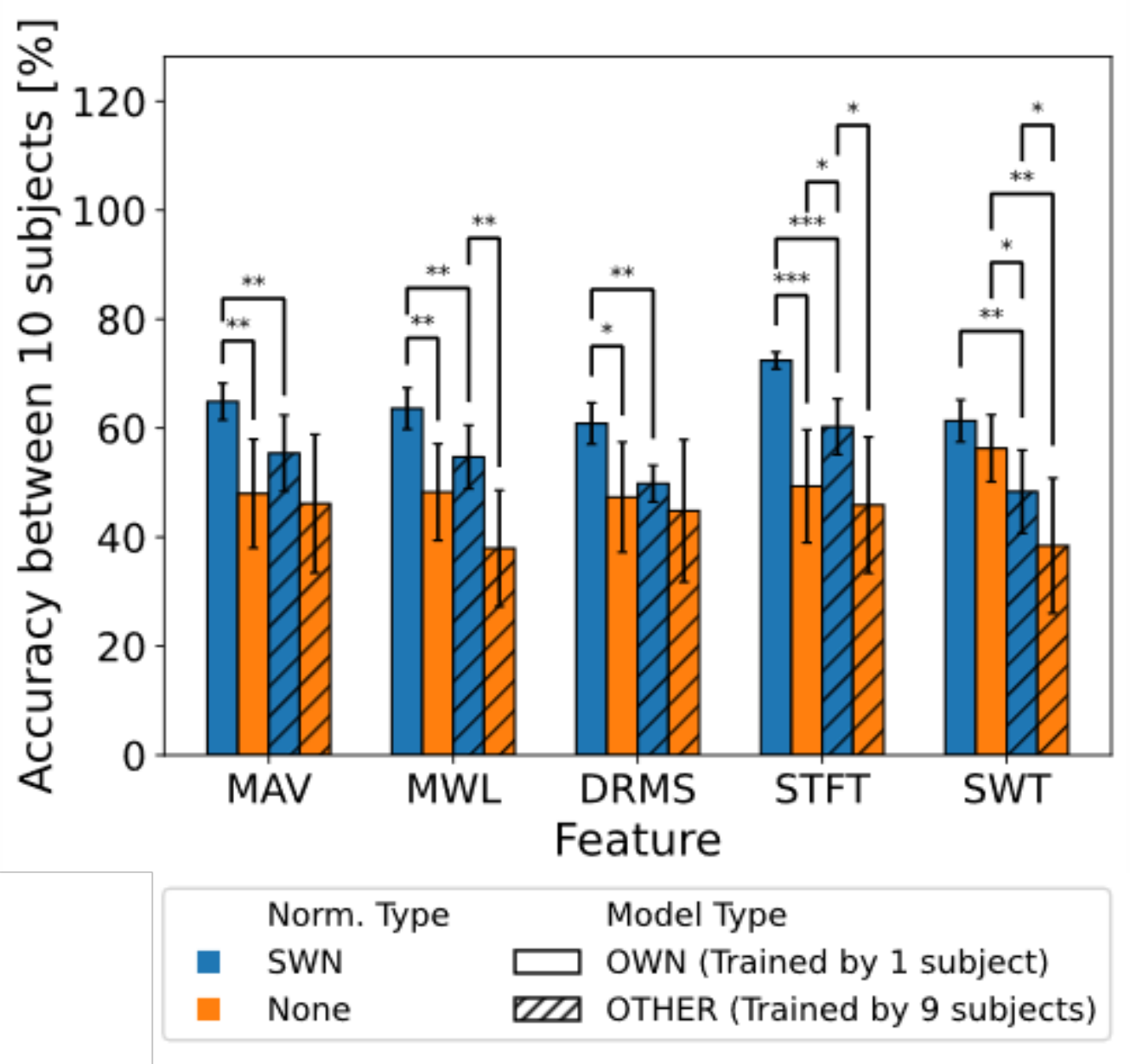}
\caption{Performance comparison between normalization methods and model types. Horizontal axis indicates feature-extraction method and vertical axis indicates accuracy using a model trained by data from other or own subjects. SWN is the result of applying normalization and None is the result of not applying normalization. Significance tests were performed between SWN\_OWN and None\_OWN, SWN\_OWN and SWN\_OTHER, None\_OTHER and SWN\_OTHER, None\_OTHER and None\_OWN, and SWN\_OTHER and None\_OTHER. The significance level was * for the P-value less than 0.05, ** for the P-value less than 0.01, and *** for the P-value less than 0.001.}
\end{figure}

\subsection{Versatility Comparison}
We investigated whether the proposed SWN method would improve versatility between users (i.e., other’s machine model would exhibit the same classification performance as one’s own model). A comparison was made between the accuracy of model OTHER with SWN applied that was used in Section III.B in Fig. 6 and model OTHER where normalization was not applied.

From Fig. 6, a comparison between SWN\_OTHER (mean accuracy: 56.5\%, mean standard deviation of accuracy: 4.7\%, blue shaded bar) and None\_OWN (mean accuracy: 49.8\%, mean standard deviation of accuracy: 9.1\%, orange bar) shows that SWN\_OTHER had a mean accuracy that was 3.9\% higher and mean standard deviation of accuracy that was 3.3\% lower. However, a comparison between SWN\_OWN (mean accuracy: 64.6\%, mean standard deviation of accuracy: 3.3\%, blue bar) and SWN\_OTHER (mean accuracy: 56.5\%, mean standard deviation of accuracy: 4.7\%, blue shaded bar) shows that SWN\_OWN had a mean accuracy that was 10.9\% higher and mean standard deviation of accuracy that was 2.5\% lower. These results show that the classification performance of the machine learning model was improved by the proposed method, but even a model that used a large amount of other’s data did not improve versatility among users to the extent that it was similar to the classification performance using one’s own data.\\
\subsection{Number of Sunjects to Train Model (OTHER)}
It was shown in Section III.B that applying the proposed SWN method could improve the classification performance of not only the model trained from one’s own data (OWN) but also the model trained from other user’s data (OTHER). Therefore, we investigated the extent to which the classification performance of the model (OTHER) could be improved by training the model by mixing the data of multiple other subjects. The number of subjects used for training the model was 1–9. The window lengths for normalization and feature extraction were changed in the range of 100 to 500 ms in 100-ms intervals, and the maximum accuracy was compared. Fig. 7 shows the results. From Fig. 7, a comparison between cases with normalization (SWN) and without normalization (None) shows that the accuracy for MAV, MWL, DRMS, and SWT for cases with normalization (SWN) hardly increased with two or more subjects used in model training. In contrast, the accuracy either monotonically increased or decreased with respect to the number of subjects for cases without normalization (None) for the five features. This implies that high classification performance can be achieved with a small number of subjects by applying SWN in cases that use other’s machine learning models. However, the accuracy is increased with respect to the number of subjects only for STFT to which SWN is applied, therefore, many subjects are required depending on the feature.\\

Next, we compared cases with either nine subjects or one subject used in the training of the machine learning model that had the highest accuracy for most of the features to investigate whether the increase in the number of subjects had a significant effect. Fig. 8 shows the results. It can be seen from Fig. 8 that SWN\_9 (mean accuracy: 53.7\%, mean standard deviation of accuracy: 5.5\%, blue bar) was the best result. Furthermore, compared to SWN\_1 (mean accuracy: 46.8\%, mean standard deviation of accuracy: 11.5\%, blue shaded bar), SWN\_9 had a mean accuracy that was 6.9\% higher and mean standard deviation of accuracy that was 5.7\% lower. Furthermore, a comparison between None\_9 (mean accuracy: 42.6\%, mean standard deviation of accuracy: 12.3\%, orange bar) and None\_1 (mean accuracy: 41.4\%, mean standard deviation of accuracy: 13.9\%, orange shaded bar) shows that None\_9 had a mean accuracy that was 1.2\% higher and a mean standard deviation of accuracy that was 1.6\% higher. These results show that the classification performance of the model (OTHER) could be improved by applying the proposed SWN method and increasing the number of subjects used for training the model (OTHER).\\



\begin{figure}[tb]
  \begin{minipage}[b]{0.45\linewidth}
    \centering
    \includegraphics[keepaspectratio, width=7cm]{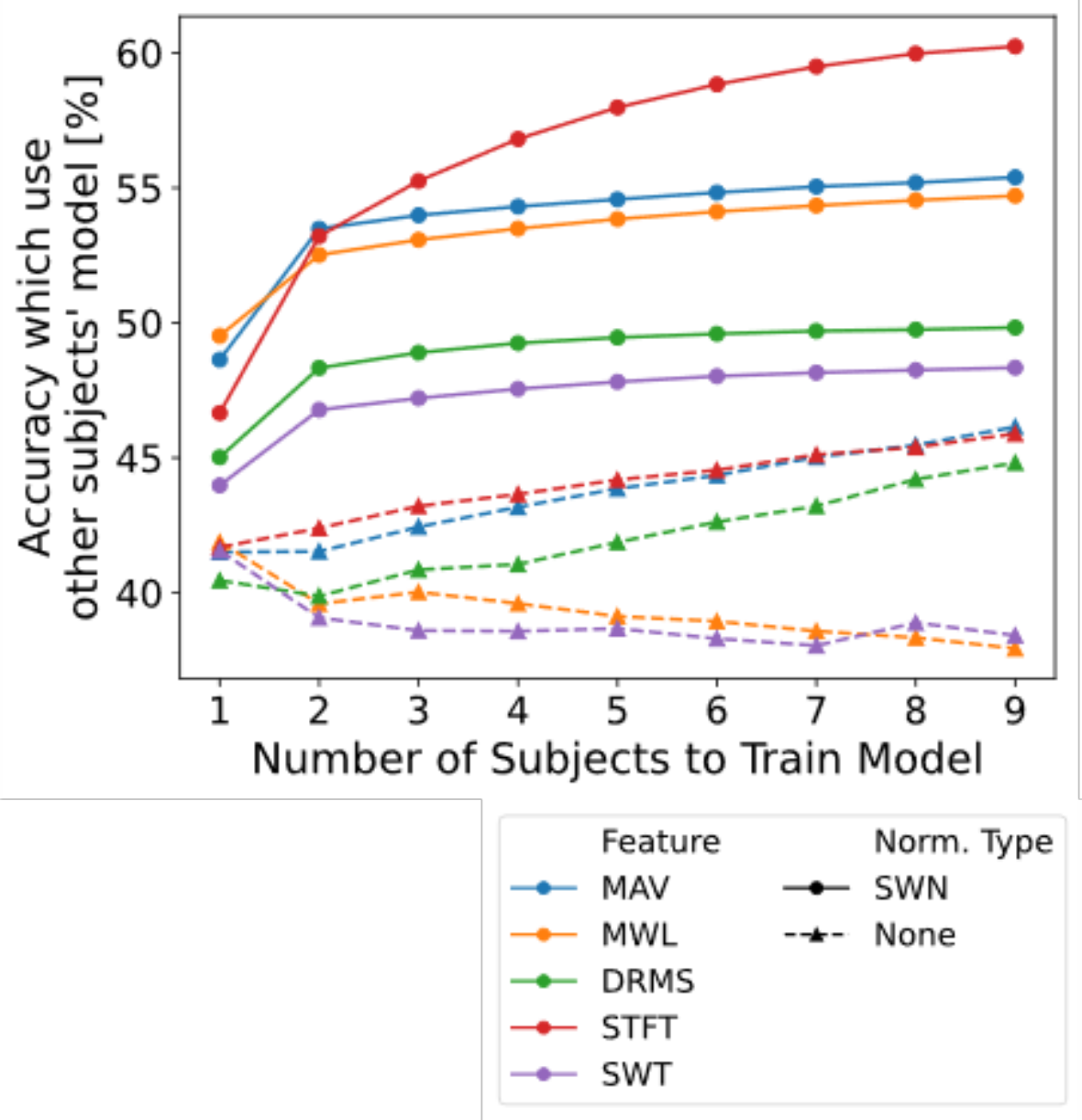}
    \caption{Effect of the number of subjects to train model (OTHER). Horizontal axis indicates number of subjects to train model (OTHER) and vertical axis indicates accuracy using a model trained by data from other subjects.}
  \end{minipage}
  \begin{minipage}[b]{0.45\linewidth}
    \centering
    \includegraphics[keepaspectratio, width=7cm]{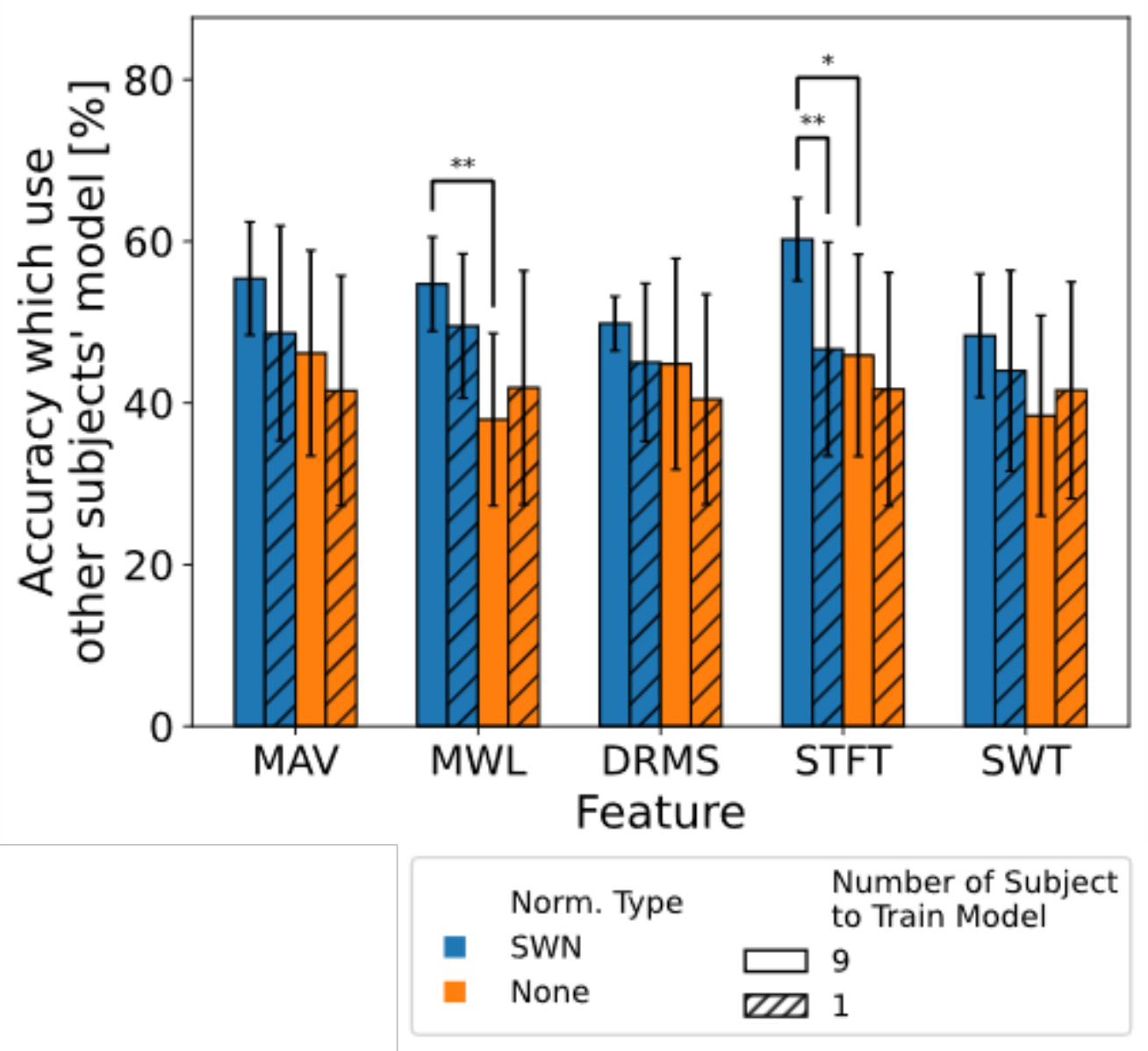}
    \caption{The performance comparison (OTHER) between normalization methods and number of subjects to train model. Horizontal axis indicates feature-extraction method and vertical axis indicates accuracy using a model trained by data from other subjects. Significance tests were performed between SWN\_9 and SWN\_1, SWN\_9 and None\_9, None\_9 and None\_1. The significance level was * for the P-value less than 0.05, ** for the P-value less than 0.01, and *** for the P-value less than 0.001.}
  \end{minipage}
\end{figure}

\subsection{Improving Performance for SWN}
It was found that the classification performance of the models (OWN and OTHER) was improved by applying the proposed SWN method in Section III.B. However, the maximum accuracy was 72.4\% (STFT in Fig. 6, blue bar), which was not a very high value. Therefore, we investigated the well-known feature extraction method in combination with the proposed SWN method.\\

First, we investigated whether the accuracy could be improved by applying the weighting window functions shown in Fig. 1 in addition to SWN. Fig. 9 shows the result of calculating the mean accuracy of the five feature types. The blue line on Fig. 9 shows the result where only the SWN was applied and the weighting window function was not applied. A comparison with no weighting window function (mean: 65.6\%) shows improved accuracy only in DownLinear (mean: 68.2\%), DownLinearStep (mean: 65.0\%), UpDownLinear (mean: 71.7\%), and UpDownLinearStep (mean: 66.1\%). In particular, UpDownLinear greatly increased the accuracy with an average of 8.1\%. These results show that UpDownLinear should be applied when applying the weighting window function.\\

Next, we investigated whether accuracy could be improved by applying the dividing window function shown in Fig. 2 in addition to SWN. Fig. 10 shows the result of calculating the mean accuracy of the five types of features. The blue line in Fig. 10 shows the result where only the SWN was applied and the dividing window function was not applied. Fig. 10 shows that the accuracy improved in all six types of dividing window functions. The accuracy was almost the same in all cases. A comparison with no dividing window function (mean: 64.6\%) shows that ED2 (mean: 73.4\%) improved by 8.8\%, ED3 (mean: 74.1\%) improved by 9.5\%, ED4 (mean: 73.4\%) improved by 8.8\%, OD2 (mean: 73.1\%) improved by 8.4\%, OD3 (mean: 73.5\%) improved by 8.9\%, and OD4 (mean: 73.7\%) improved by 9.0\%. These results show that an appropriate window function should be selected in consideration of reducing the amount of calculation when applying the dividing window function. This implies that ED2 is appropriate for time-domain features such as MAV and MWL that could be calculated with short-length signals, and OD2 is appropriate for time-frequency domain features such as STFT and SWT that require a somewhat long signal.\\

We were able to improve the model classification performance by applying the two window function types of weighting window functions and dividing window functions. However, the maximum mean accuracy was 74.1\% (ED3 in Fig. 10), which is still a low value. Therefore, further improvements to classification performance are needed.\\



\begin{figure}[tb]
  \begin{minipage}[b]{0.45\linewidth}
    \centering
    \includegraphics[keepaspectratio, width=7cm]{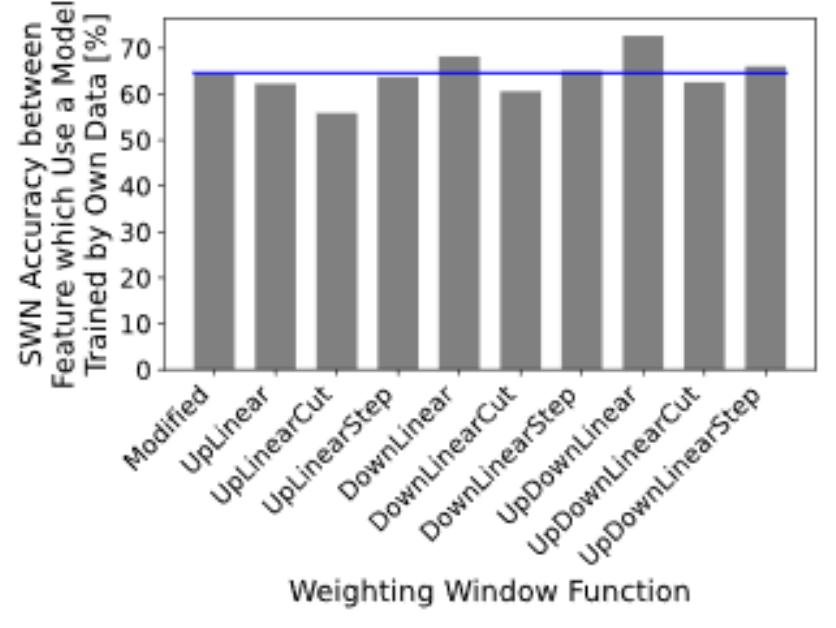}
    \caption{Performance comparison between weighting window functions. The blue line indicates the SWN accuracy without weighting window function. Horizontal axis indicates weighting window functions in Figure 1 and vertical axis indicates accuracy between 5 types of features using a model trained by data from own subject.}
  \end{minipage}
  \begin{minipage}[b]{0.45\linewidth}
    \centering
    \includegraphics[keepaspectratio, width=7cm]{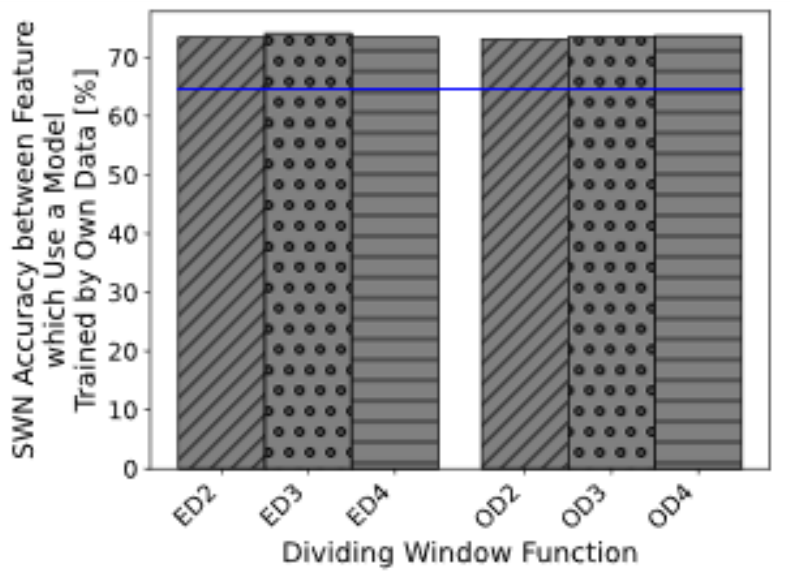}
    \caption{Performance comparison between dividing window functions. The blue line indicates the SWN accuracy without dividing window function. Horizontal axis indicates dividing window function in Figure 2 and vertical axis indicates accuracy between 5 types of features using a model trained by data from own subject. ED2 is Equal Division which the number of the dividing windows is two, OD2 is Overlap Division which the number of the dividing windows is two.}
  \end{minipage}
\end{figure}

\section{Disscussion}
In this study, we proposed a new normalization method, SWN, to improve the classification performance of machine learning models. We succeeded in increasing mean model accuracy from 49.8\% to 64.6\%, an increase of 15.0\%, by applying the SWN (blue and orange bar with no line in Fig. 6). Furthermore, the mean standard deviation of accuracy decreased from 9.1\% to 3.3\%, a decrease of 5.8\%. The results show the effectiveness of the proposed method.\\

In this section, we analyze the factors that improve the model classification performance by the proposed method (Section IV.A) and discuss the feasibility of real-time prediction (Section IV.B).\\
\subsection{Analysis of SWN}
Section III.B showed the effectiveness of SWN in improving the classification performance of the models. We investigated the effect of dividing with the standard deviation of EMG, which was thought to have led to the improvement of the classification performance of machine learning models and is a feature of SWN. We conducted an analysis by visualization of the standard deviation of EMG vs. feature distribution. The visualization of the S.D. of EMG vs. the feature distribution was conducted by drawing a confidence ellipse with a standard deviation of 2. The “\href{https://matplotlib.org/devdocs/gallery/statistics/confidence_ellipse.html}{confidence\_ellipse}” function of matplotlib in Python was used for implementation. Fig. 11 shows an example of the results of treating MAV as a representative of the features. Fig. 11 shows that the S.D. of EMG-MAV distribution in the case with normalization (SWN) had a weakly negative or no correlation, whereas the distribution in the case without normalization (None) had a strongly positive correlation. \\

We used the results obtained in Fig. 11 as a basis for conducting a correlation analysis of cases with normalization (SWN) and without normalization (None). The representative feature was MAV, which was the same as in Fig. 10. Fig. 12 shows the results of calculating the correlation coefficient of the S.D. of EMG vs. MAV for each channel and subject and taking the mean value. It can be seen from Fig. 12 that the S.D. of EMG and MAV had a weakly negative correlation for cases with normalization (SWN) and a strongly positive correlation for cases without normalization (None). These results imply that one of the factors that improved the classification performance of the machine learning model was the reduction of the influence of the standard deviation on the features by the proposed SWN method.\\

\begin{figure}[tb]
  \begin{minipage}[b]{0.45\linewidth}
    \centering
    \includegraphics[keepaspectratio, width=7cm]{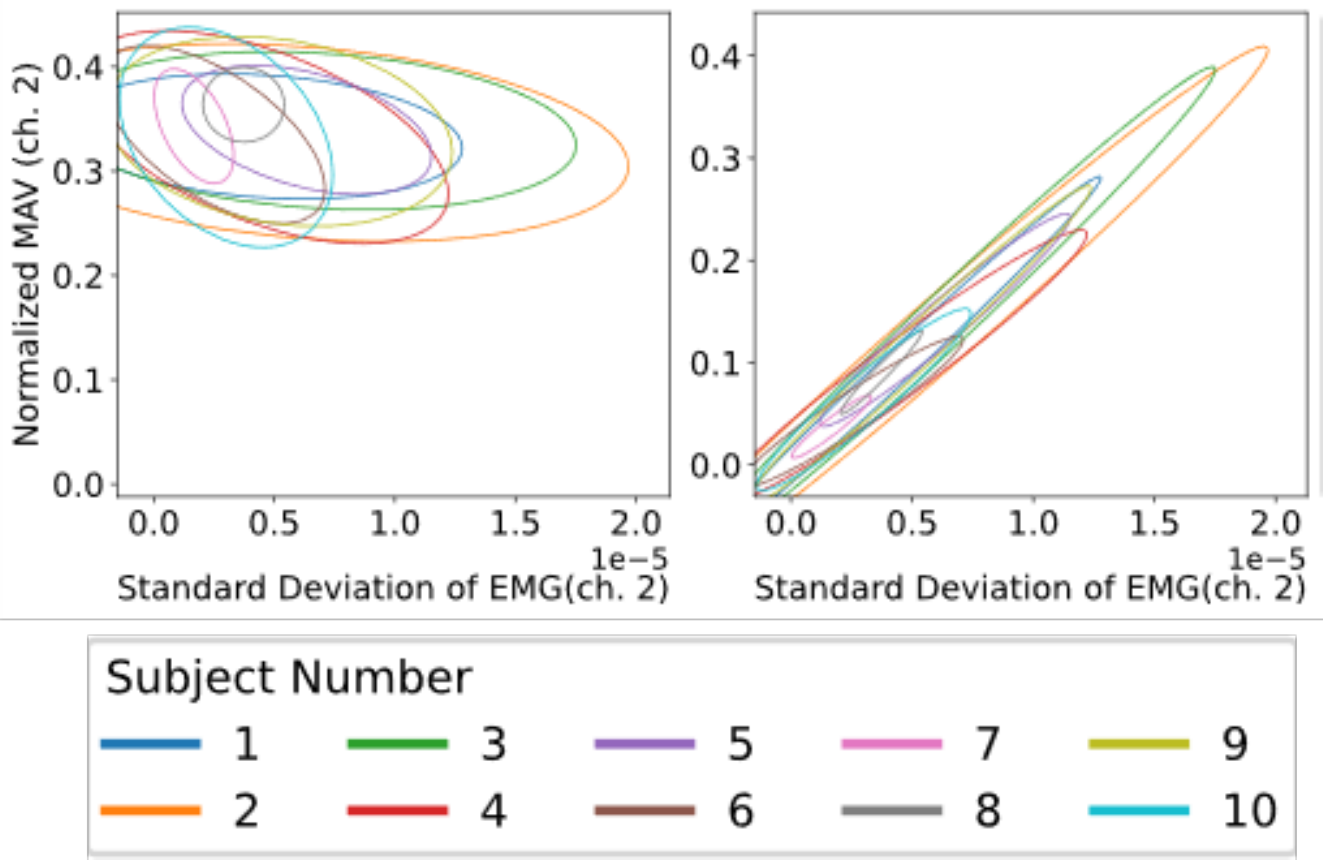}
    \caption{Example of the distribution analysis between standard deviation of EMG and feature each subject on the channel 2 (MAV). (A) Normalization (SWN), (B) No Normalization (None). Both MAV of SWN and None are normalized by maximum data in the 10 subjects.}
  \end{minipage}
  \begin{minipage}[b]{0.45\linewidth}
    \centering
    \includegraphics[keepaspectratio, width=7cm]{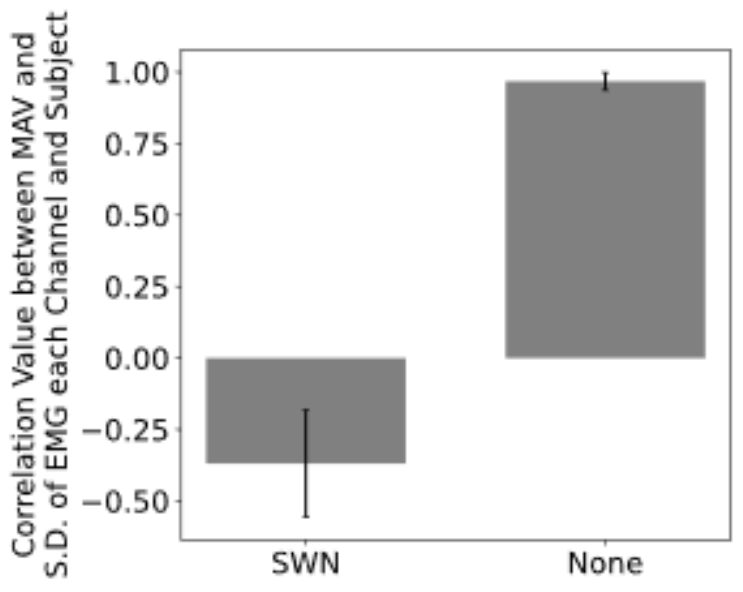}
    \caption{Correlation comparison between standard deviation of EMG and MAV. SWN means with normalization and None means without normalization.}
  \end{minipage}
\end{figure}

\subsection{Comparison of the Calculation Time}
It was shown in Section III.B that SWN was effective in improving the classification performance of the machine learning model. However, it is still unknown whether this can satisfy the required execution speed for real-time processing. Therefore, the pre-processing and normalization shown in Section II.C were executed at intervals of 20 ms (50 Hz), and the mean execution time was compared between cases with normalization (SWN) and without normalization (None). The execution environment was as follows: Intel(R)Core (TM) i7-9700K CPU @ 3.60 GHz 3.60 GHz, Python 3.8.12. Fig. 13 shows the results. A comparison of the mean execution times shows that SWN (mean: 409 µs) was longer by 76 µs than None (mean: 333 µs). The normalization rate in 409 µs was 18.6\%, a minimal effect on real-time processing. These results imply that the proposed SWN method can be implemented in real-time.\\
\begin{figure}[tb]
\centering
\includegraphics[keepaspectratio, width=7cm]{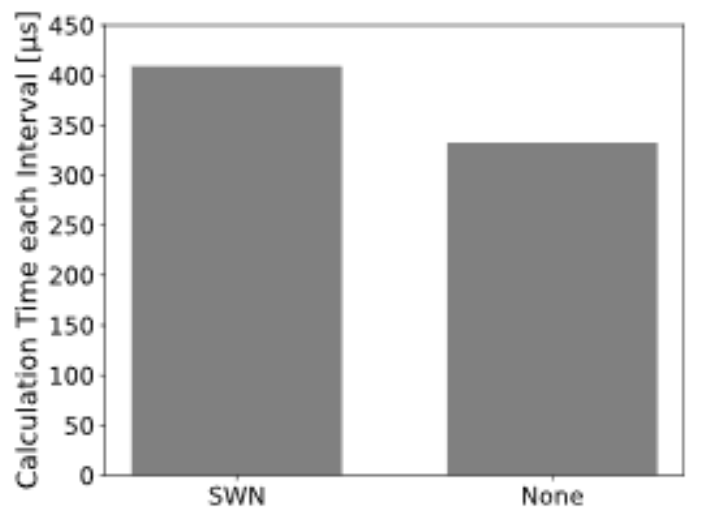}
\caption{Comparison of the calculation time. SWN means normalization and None means no normalization.}
\end{figure}
\section{Conclusion}
In this paper, we proposed a normalization method that used the sliding-window and z-score (SWN) to improve the classification performance of devices using EMG. Applying SWN improved the mean accuracy by 15.0\% compared to the case without normalization. Even when a machine learning model that was trained with other’s data was used, the mean accuracy improved by 11.1\% compared to the case without normalization. These results show that the classification performance of the machine learning model could be improved by the proposed method (SWN). Results of investigating the relationship between the standard deviation and features also show that applying the SWN changed the correlation between the standard deviation and features from that of a strongly positive one to a weakly negative one. This was assumed to be one of the factors that improved the classification performance of machine learning models.\\

Furthermore, the maximum mean accuracy was low at 64.6\%, so we proposed the weighting window function and dividing window function as methods for improving accuracy. The maximum mean accuracy was 73.7\%, for an increase of 9.0\%.\\

Future issues include applications not only to classification models but also regression models. In this paper, the proposed method was applied only to the classification model. Therefore, the proposed method needs to be applied to a regression model that predicts kinematic parameters such as the joint angle and joint angular velocity. \\


\end{document}